\documentstyle[psfig,amssymb]{mn}
\title[The globular cluster system of the young elliptical NGC\,6702] 
{The globular cluster system of the young elliptical NGC\,6702}

\author[Georgakakis, Forbes \& Brodie] {
  Antonis E. Georgakakis$^{1}$\thanks{age@star.sr.bham.ac.uk},
  Duncan A. Forbes$^{1,2}$\thanks{forbes@star.sr.bham.ac.uk}, 
  Jean P. Brodie$^{3}$\thanks{brodie@ucolick.org}\\ \\
  $^1$ School of Physics and Astronomy, University of Birmingham,
  Edgbaston, B15 2TT, UK\\
  $^2$ Astrophysics \& Supercomputing, Swinburne University,
  Hawthorn, VIC 3122, Australia\\
  $^3$ Lick Observatory, University of California, Santa Cruz, CA 95064, USA\\
}
\begin{document}
\maketitle

\begin{abstract}
We study the globular cluster (GC) system of the dust-lane elliptical
galaxy NGC\,6702, using $B$, $V$ and  $I$-band photometric observations
carried out at the Keck telescope. This galaxy has a spectroscopic age
of $\approx2$\,Gyrs suggesting recent star-formation. We find strong
evidence for a bimodal GC  colour distribution, with the blue peak
having a colour similar to that of the Galactic halo GCs. Assuming
that the blue GCs are indeed old and metal--poor, we estimate an age
of 2--5\,Gyrs and supersolar metallicity for the red GC
subpopulation. Despite the large uncertainties, this is in reasonable
agreement with the spectroscopic galaxy age. Additionally, we
estimate a specific frequency of $S_{N}=2.3\pm1.1$ for NGC\,6702. We
predict that passive evolution of NGC\,6702 will further increase its  
specific frequency to  $S_{N}\approx2.7$ within 10\,Gyrs, in closer
agreement to that of  typical present-day ellipticals.  We also discuss
evidence that the merger/accretion event that took place a few Gyrs ago
involved a high gas fraction. 
\end{abstract}


\section{Introduction}\label{sec_intro}

The `merger' hypothesis, postulating that elliptical galaxies form by
spiral galaxy mergers, was first proposed by Toomre \& Toomre
(1972). This scenario has gained new impetus over the last few years by
both numerical simulations (e.g. Barnes 1992; Hernquist 1992; Mihos \&
Hernquist 1996) and observational studies, suggesting that the properties
of merger remnants are consistent with those of ellipticals (e.g. Toomre
1977;  Schweizer et al. 1990;  Casoli et al. 1991; Schweizer \& Seitzer
1992; Hibbard \& van Gorkom 1996; Georgakakis, Forbes \& Norris 2000).  

However, one of the long standing drawbacks of the `merger' hypothesis is
the excess globular clusters (GCs) per unit starlight around ellipticals
compared to spirals (van den Bergh 1984). To overcome this problem
Schweizer (1987) and Ashman \& Zepf (1992) suggested that the starburst
activity usually experienced during gas-rich mergers will form new GCs and
thus increase their number compared to their progenitors.   Indeed, there
is accumulating observational evidence supporting this scenario in on-going
mergers (e.g. Lutz 1991; Whitmore et al. 1993; Whitmore \& Schweizer 1995;
Schweizer 1996; Miller et al. 1997; Zepf et al. 1999; Forbes \& Hau 2000). 
However, the situation in ellitpical galaxies (the eventual product of
mergers) is less clear. Although many ellipticals have been observed to
have bimodal colour distributions, when examined in detail their GC
properties do not match those from the merger picture (Forbes et
al. 1997). Outstanding questions include whether the newly formed GCs
survive and whether the final GC system resembles that of typical (old) 
ellipticals. 
Probing the GC properties in intermediate age merger-remnants (1--5\,Gyrs)
may be of key importance for addressing these issues. Indeed, such
young ellipticals may provide the link between on-going mergers and
relaxed ellipticals.   

For example, a sample of merger--remnants spanning a range of ages
will allow study of the chronological development of the system's
specific frequency (measuring the number of globular clusters per unit
starlight) in comparison with that of evolved ellipticals. Young
ellipticals are also expected to have two distinct GC populations 
with different colour and magnitude distributions. By exploiting these
differences one can attempt to age-date the newly formed GCs, which in turn
can be compared to the time since the gaseous merger. Indeed, studies of
the  dynamically young proto-ellipticals NGC\,1700 and NGC\,3610  have
shown the presence of bimodality in their colour distributions (Whitmore et
al. 1997; Brown et al. 2000). Also the age of the new GCs, although
model dependent, is found to be similar to estimates for the galaxy age
based on spectroscopic or dynamical arguments.

In this paper we study the globular cluster system of the elliptical galaxy
NGC\,6702. Lauer (1985) found a dust lane in the centre of NGC\,6702
indicating a recent gaseous merger event. This galaxy is also relatively
luminous  at far-infrared wavelengths compared to other elliptical
galaxies, with  $L_{FIR}=1.3\times10^{10}\,\mathrm{L_{\odot}}$  (Wiklind et
al. 1995).  Terlevich \& Forbes (2000) have estimated a spectroscopic age
and metallicity of about 2\,Gyrs and $\mathrm{[Fe/H]=+0.5}$  respectively for  
NGC\,6702 suggesting that it is a young elliptical.
The main observational properties of NGC\,6702 are given in Table
\ref{tab_0}. 

Additionally, according to the NED database the nearest galaxy projected
on the sky that is either brighter or no more than 2 magnitudes
fainter, than NGC\,6702 is  NGC\,6703 at a projected distance of
10\,arcmin. However, as NGC\,6702 has a recession velocity of
$4727\,\mathrm{km\,s^{-1}}$ and NGC\,6703 has $V=2365\mathrm{\,km\,s^{-1}}$
they are actually separated by some 30\,Mpc  assuming no peculiar motions.
The nearest galaxy with a similar recession velocity is NGC\,6711 
($V=4671\mathrm{\,km\,s^{-1}}$) with a separation of $\sim120$\,arcmin or
over 2\,Mpc. Other galaxies listed by NED are more than 2 magnitudes fainter
than NGC\,6702. Thus NGC\,6702 is an example of an isolated elliptical
galaxy with no comparable galaxies within 2\,Mpc.  Such galaxies are fairly
rare and hence their evolutionary history is of some interest. 
In particular, ellipticals in low density environments are interesting
within the hierarchical merger framework as they are expected to have
assembled much more recently than their cluster counterparts (Kauffmann \&
Charlot 1998). Any merger would be expected to have occurred in the more
recent past and involved progenitors that have largely processed gas into
stars.  

In section \ref{sec_obs} the observations, data reduction and  the 
isophotal properties of NGC\,6702 are presented. Section \ref{sec_sel} 
describes the candidate GC selection, while section \ref{sec_res} 
presents our results which are discussed in section
\ref{sec_disc}. Finally,  section \ref{sec_conc} summarises our
conclusions. Throughout this  paper we adopt
$H_0=75\,\mathrm{km\,s^{-1}\,Mpc^{-1}}$.  

\begin{table*} 
\footnotesize 
\begin{center} 
\begin{tabular}{cccccccccc} 
\hline 
   Name  &  RA (J2000) & DEC (J2000) & l & b & type & heliocentric& distance 
          & $B_T^0$ & $V_T^0$ \\        
          &             &             &   &   & (RC3)& velocity ($\mathrm{\,km\,s^{-1}}$)  &  modulus  
          &  (mag)    &  (mag)\\
 NGC\,6702   
 &  $18\mathrm{^h}$ $46\mathrm{^m}$ $57.6\mathrm{^s}$
 &  $+45\mathrm{^\circ}$ $42\mathrm{^\prime}$ $20\mathrm{^{\prime\prime}}$  
 &  $75.03\mathrm{^\circ}$
 &  $19.79\mathrm{^\circ}$
 &  E\,3
 &  4727
 &  34.00
 &  13.04 
 &  12.15 \\ \hline
\end{tabular} 
\end{center} 
\caption{Observational properties of NGC\,6702 obtained from NED. The
distance modulus is estimated assuming
$H_0=75\,\mathrm{km\,s^{-1}\,Mpc^{-1}}$.}\label{tab_0}  
\normalsize  
\end{table*} 

\section{Observations and data reduction}\label{sec_obs}

\subsection{Description of the data}\label{sec_data}

Broad-band imaging of NGC\,6702 in the  $B$, $V$ and $I$ filters was
carried out at the Keck-II telescope on 1999 August 17, using the the Low
Resolution Imaging Spectrometer (LRIS; Oke et al. 1995). The LRIS
instrument, equipped with a TEK $2048\times2048$ CCD, is mounted on the
Cassegrain focus providing a
$0.215\,\mathrm{arcsec\,pixel^{-1}}$ imaging scale and a
$6^{\prime}\times8^{\prime}$ field--of--view. The total exposure time in
the $B$, $V$ and  $I$ filters was 3600, 1800, and 1200\,sec respectively,
segregated into three separate integrations of 1200, 600 and 400\,sec
respectively. The seeing at the time of the observations varied between
0.8 to 1\,arcsec. 

\begin{figure} 
\centerline{\psfig{figure=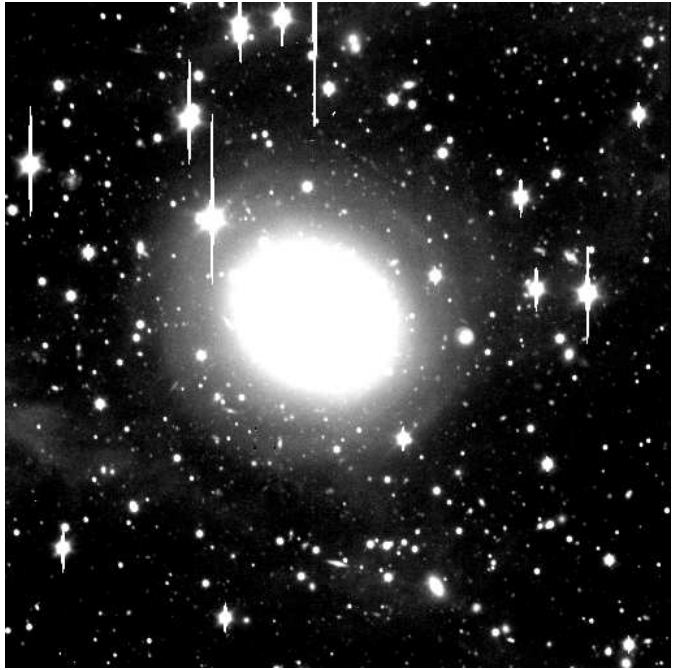,width=0.5\textwidth,angle=0}} 
 \caption
{$V$-band image of NGC\,6702 obtained at the Keck telescope.}\label{fig_image}
\end{figure}

The data were reduced following standard procedures, using IRAF tasks. 
The reduced images are found to be flat to better than $\sim2\%$. 
The final $V$-band image of NGC\,6702 is shown in Figure
\ref{fig_image}. Photometric calibration was performed using standard stars
from Landolt (1992). The photometric accuracy, estimated  using these 
standard stars, is $\pm0.02$\,mag in all three bands. We adopt a Galactic
extinction in the NGC\,6702 direction of $E(B-V)=0.06$, taken from Burstein
\& Heiles (1984). Assuming a standard  reddening curve (Savage \& Mathis
1979), we correct each magnitude for Galactic extinction using the values 
$A_{B}=0.25$,  $A_{V}=0.19$ and $A_{I}=0.09$\,mag.  

\subsection{Galaxy surface brightness profile}

\begin{figure} 
\centerline{\psfig{figure=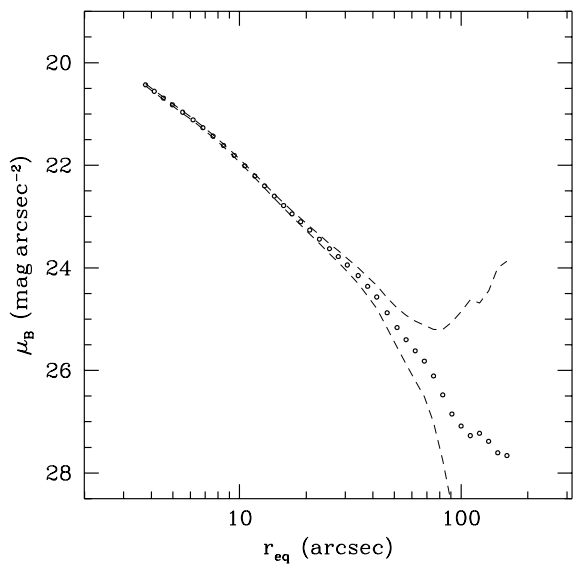,width=0.5\textwidth,angle=0}} 
 \caption
{$B$-band surface brightness profile of NGC\,6702. The dashed lines are the 
$1\sigma$ uncertainty envelopes. We define the equivalent radius
$r_{eq}=a\,\sqrt{1-\epsilon}$, where $a$ and $\epsilon$ are the semi-major 
axis and ellipticity respectively.}\label{fig_prof}
\end{figure}

The $B$-band surface brightness profile, $\mu_B$, of NGC\,6702 is plotted in 
Figure \ref{fig_prof}. This has been determined by fitting ellipses to the
galaxy intensity profile using the ISOPHOTE task within STSDAS. 
During the ellipse fitting process  the centre of
the galaxy was kept fixed and a $3\,\sigma$ clipping algorithm was
employed. Also, bright objects as well as the central saturated parts of 
the galaxy, were masked out. The position angle (PA) and ellipticity
($\epsilon$) were fitted at each radius, until S/N constraints terminated
the fitting process at low surface brightness. For NGC\,6702 this
occurs at an equivalent radius $r_{eq}\approx50^{\prime\prime}$
($r_{eq}=a\,\sqrt{1-\epsilon}$, where $a$ is the semi-major axis). Beyond
that radius the position angle and ellipticity were kept constant. The sky
level was determined using the method described by Goudfrooij et
al. (1994).

Although the isophotes of  elliptical galaxies are  well  approximated by
ellipses, deviations at the few per cent level are common. Such deviations
are often quantified by the 4th sine and cosine amplitudes (S4 and C4  
respectively) of the Fourier transform of the galaxy intensity at a given
radius. The standard interpretation of the residual C4 terms is that they
indicate either the presence of a weak underlying disk if positive or 
boxiness if negative. However, it has been emphasised that both C4 and S4
terms should be considered when estimating the isophotal shape of galaxies
(Franx et al. 1989; Peletier et al. 1990; Goodfrooij et al. 1994).  For
example, the presence of a disk, bar or ring with a significant position
angle twist relative to the elliptical galaxy semi-major axis is expected to 
give both positive and negative C4 and/or S4 terms, depending on its
orientation. Consequently, in the case of such projection effects we expect
S4 terms to be significant.

\begin{figure} 
\centerline{\psfig{figure=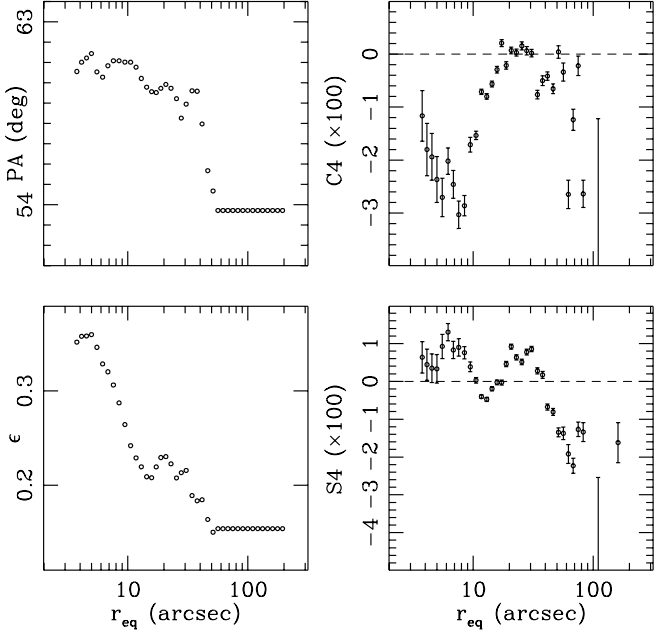,width=0.5\textwidth,angle=0}} 
 \caption
{Radial profile in the $B$-band of the ellipticity ($\epsilon$), position
angle (PA), 4th cosine term (C4)  and 4th sine term (S4), estimated by
fitting ellipses to the galaxy light profile. At radii larger than
$\approx50^{\prime\prime}$ both the ellipticity and the position angle are
kept constant during the ellipse fitting routine, due to 
S/N constraints.}\label{fig_prof1} 
\end{figure}

Figure \ref{fig_prof1} plots the ellipticity, position angle as well as C4
and S4 terms  as a function of radius for NGC\,6702 in the
$B$-band. Similar trends are apparent for the $V$ and $I$-bands. 
The C4 
term is negative at small galactocentric distances, indicating
boxiness, i.e. excess light at 45$^{o}$ relative to the semi-major
axis. This is partly due to the presence of the dust lane in the core of
NGC\,6702 (Lauer 1985) lying perpendicular to the  semi-major axis and
extending 
out to $\approx20^{\prime\prime}$. The S4 term is also non-zero within the same
range of $r_{eq}$, albeit with a smaller amplitude. Masking out the dust
lane  reduces the amplitude of both the C4 and S4 terms. These terms are
also non-zero (S4 positive and C4 negative) at
$r_{eq}\approx40^{\prime\prime}$ indicating boxiness. At larger distances 
($r_{eq}>60^{\prime\prime}$) both the C4 and S4 deviate significantly from 
the elliptical symmetry, but this is likely to be an artifact arising from
the low S/N at these radii. 

\subsection{Source extraction and completeness limits}

To detect sources superimposed on NGC\,6702 we subtract the galaxy
light profile using a median filter technique. Firstly, the reduced image  
for each filter is smoothed using  a median rectangular filter with a size 
of 40 pixels. The smoothed image is then subtracted from the original
frame and the resulting image is subsequently used for source extraction
and photometry.    

The sources in the reduced frames were extracted using the
SExtractor package (version 2.1.0; Bertin \& Arnouts 1996). The main input  
parameters are the detection threshold, given as a multiple of the sky
variance ($\sigma_{Sky}$) and the minimum area in pixels for an object to
be extracted. After considerable experimentation, we adopted a threshold of 
2.5$\times \sigma_{Sky}$ and a minimum area of 10--15\,pixels
($\approx$0.5--0.7\,arcsec$^{2}$).  This choice of values minimises the
number of spurious detections, while ensuring that faint objects are
successfully extracted. A total of 1455, 1456 and 1494 sources were
extracted form the $B$, $V$ and $I$ frames respectively. The photometric
catalogues generated for each waveband were then matched, resulting in a
total of 939 common detections in all three filters, $B$, $V$ and $I$. This
source list is referred to as the  matched catalogue. Using this catalogue
we performed photometry using the PHOT task. After a curve--of--growth
analysis on several sources an aperture of 8 pixels radius was 
employed with the background estimated from an annulus of 15 to 20 pixels
around each source. For unresolved sources the 8 pixels radius aperture
magnitude is close to total. 

\begin{figure} 
\centerline{\psfig{figure=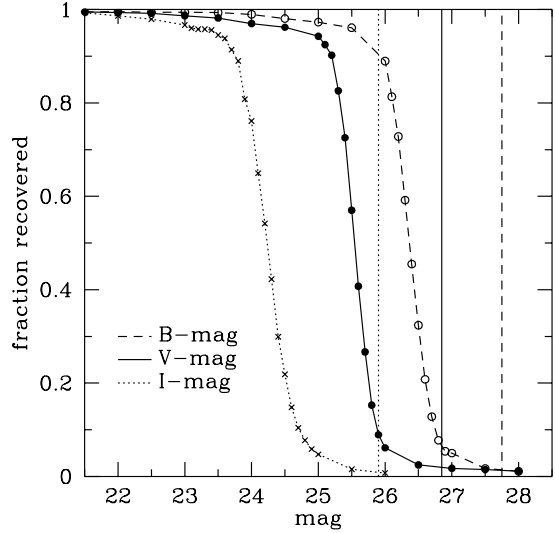,width=0.5\textwidth,angle=0}} 
 \caption
{Fraction of successfully recovered point sources as a function of 
magnitude in the $B$ (dashed line), $V$ (continuous line) and $I$ (dotted
line) filters. The vertical lines represent the peak magnitude of the Milky
Way GC luminosity function in different wavebands (Ashman \& Conti 1995;
$M_{B,peak}=-6.50$, $M_{V,peak}=-7.33$, $M_{I,peak}=-7.95$), shifted to the
distance of NGC\,6702.}\label{fig_complete}   
\end{figure}

The completeness limit in a given filter is estimated by adding artificial
point sources of different magnitudes to the original image, using
MKOBJECTS task. Then we attempt to recover the
artificial objects using  SExtractor package, with the same parameter
settings as with those adopted for the real data. A total of 1500 point
sources were added to a given filter frame. To avoid
overcrowding the frame, artificial sources are added in groups of 150 and 
the process is repeated 10 times. The results are  presented in Figure
\ref{fig_complete}, plotting the fraction of recovered objects as a
function of magnitude in the $B$, $V$ and $I$ filters respectively. The
$80\%$ completeness limits in the $B$, $V$ and $I$ filters are found to
be $\approx26.1$, $\approx25.3$ and $\approx23.9$\,mag respectively. The
mean photometric error at these magnitude limits is $\approx0.1$\,mag for
all three filters. We have also investigated variations of the
completeness limits due to the increased background close to the galaxy
centre. A method similar to that described above was adopted, with
artificial point sources being added within successive elliptical rings
centred  on NGC\,6702 and having the same mean ellipticity and position
angle as the galaxy ($\epsilon=0.2$, PA=145$^{\circ}$). The effect of
galaxy background was found to be small except for
$r_{eq}\lesssim25^{\prime\prime}$. However, this is close to the saturated
region of the CCD images which is masked out in our analysis. Therefore,
the lower sensitivity close to the galaxy center will not affect the
results from this study.

\section{Selection of globular cluster candidates}\label{sec_sel}

\begin{figure} 
\centerline{\psfig{figure=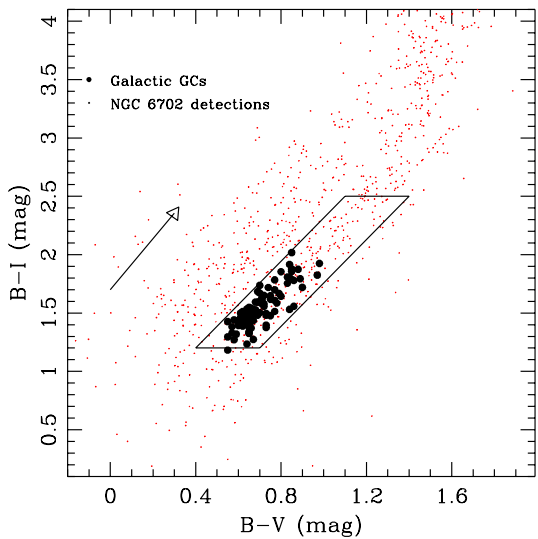,width=0.5\textwidth,angle=0}} 
 \caption
{$B-I$, $B-V$ colour-colour diagram for the sources detected in all three
filters, $B$, $V$ and $I$ (dots). The Milky Way globular clusters are 
shown as filled circles. The lines delimit the region of the parameter
space that globular clusters are expected to occupy (see text for
details). The arrow indicates the direction the $B-V$ and $B-I$
colours redden for a single burst stellar population model of constant
age and increasing metallicity (Worthey 1994).}\label{fig_bv_bi}   
\end{figure}

\begin{figure} 
\centerline{\psfig{figure=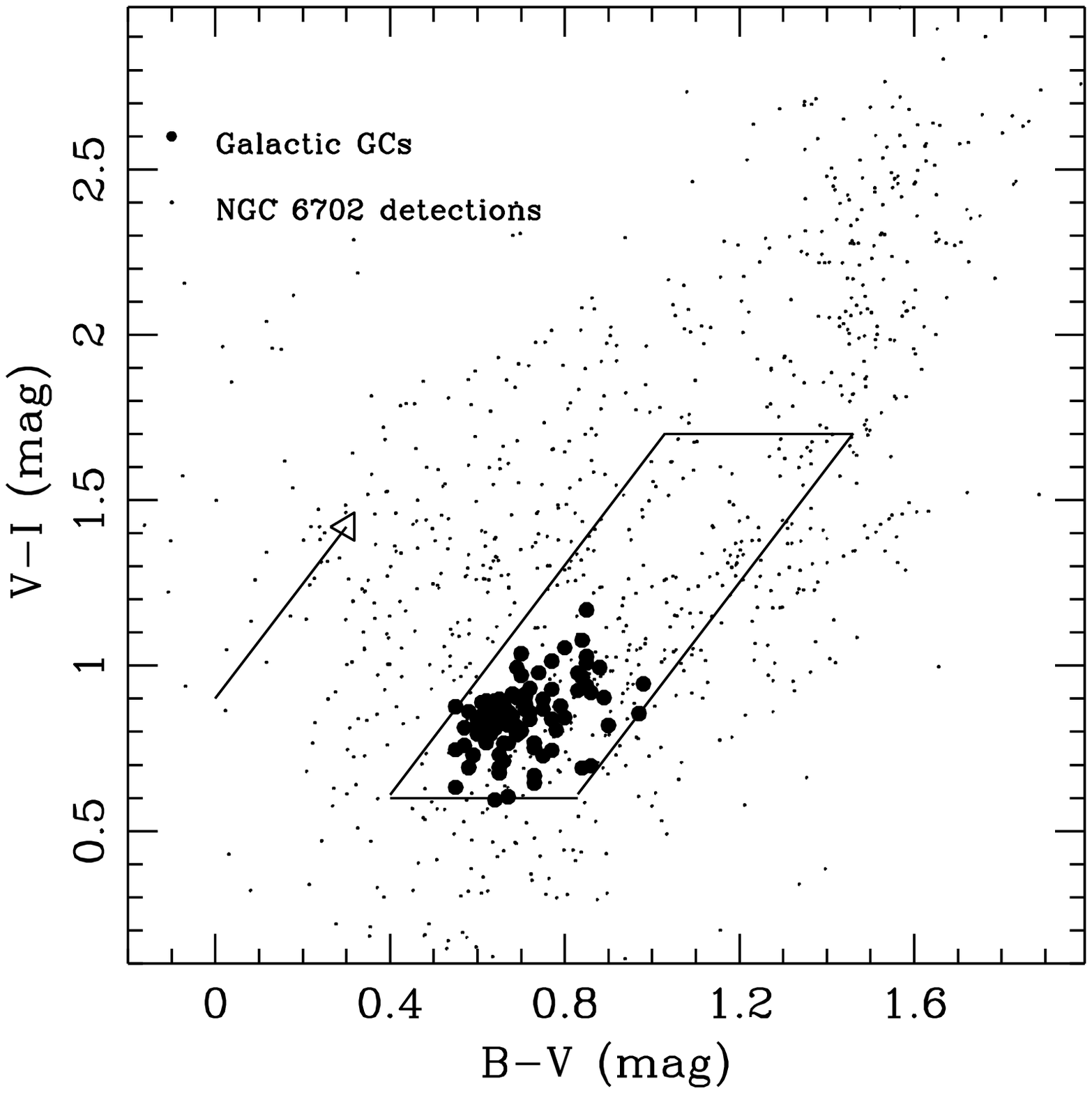,width=0.5\textwidth,angle=0}} 
 \caption
{$V-I$, $B-V$ colour-colour diagram for the sources detected in all three
filters, $B$, $V$ and $I$ (dots). The Milky Way globular clusters are
shown as filled circles. The lines delimit the region of the parameter
space that globular clusters are expected to occupy (see text for
details). The arrow indicates the direction the $B-V$ and $V-I$
colours redden for a single burst stellar population model of constant
age and increasing metallicity (Worthey 1994).}\label{fig_bv_vi}   
\end{figure}

Globular cluster candidates are selected from the matched 
catalogue on the basis of their expected range in colour and magnitude. 
In particular,
the full range of possible globular cluster metallicities is
$-2.5<\mathrm{[Fe/H]}<+1.0$. Using the colour--metallicity 
relation of Couture et al. (1990) and Kissler-Patig et al. (1998), the
above mentioned metallicity interval corresponds to $1.2<B-I<2.5$ and
$0.6<V-I<1.7$. Figures \ref{fig_bv_bi} and \ref{fig_bv_vi}
plot the $B-I$ and $V-I$ colours against $B-V$ respectively, for the
objects in the matched catalogue. The horizontal
lines in these figures represent the above mentioned $B-I$ and $V-I$ colour 
cutoffs. Also shown are the regions of the parameter space 
occupied by the Galactic GCs (Harris 1996; corrected for Galactic
reddening). The accepted range of $B-V$ colour for globular cluster 
selection is chosen  to span the $B-V$ colour range of the Galactic
GCs. Additionally, using the population synthesis models of Worthey 
(1994) we find that the  $B-I$, $V-I$ and $B-V$ colours of a stellar
population of a given age get redder with increasing metallicity in the
direction indicated by the arrows in Figures  \ref{fig_bv_bi} and
\ref{fig_bv_vi}. 

\begin{figure} 
\centerline{\psfig{figure=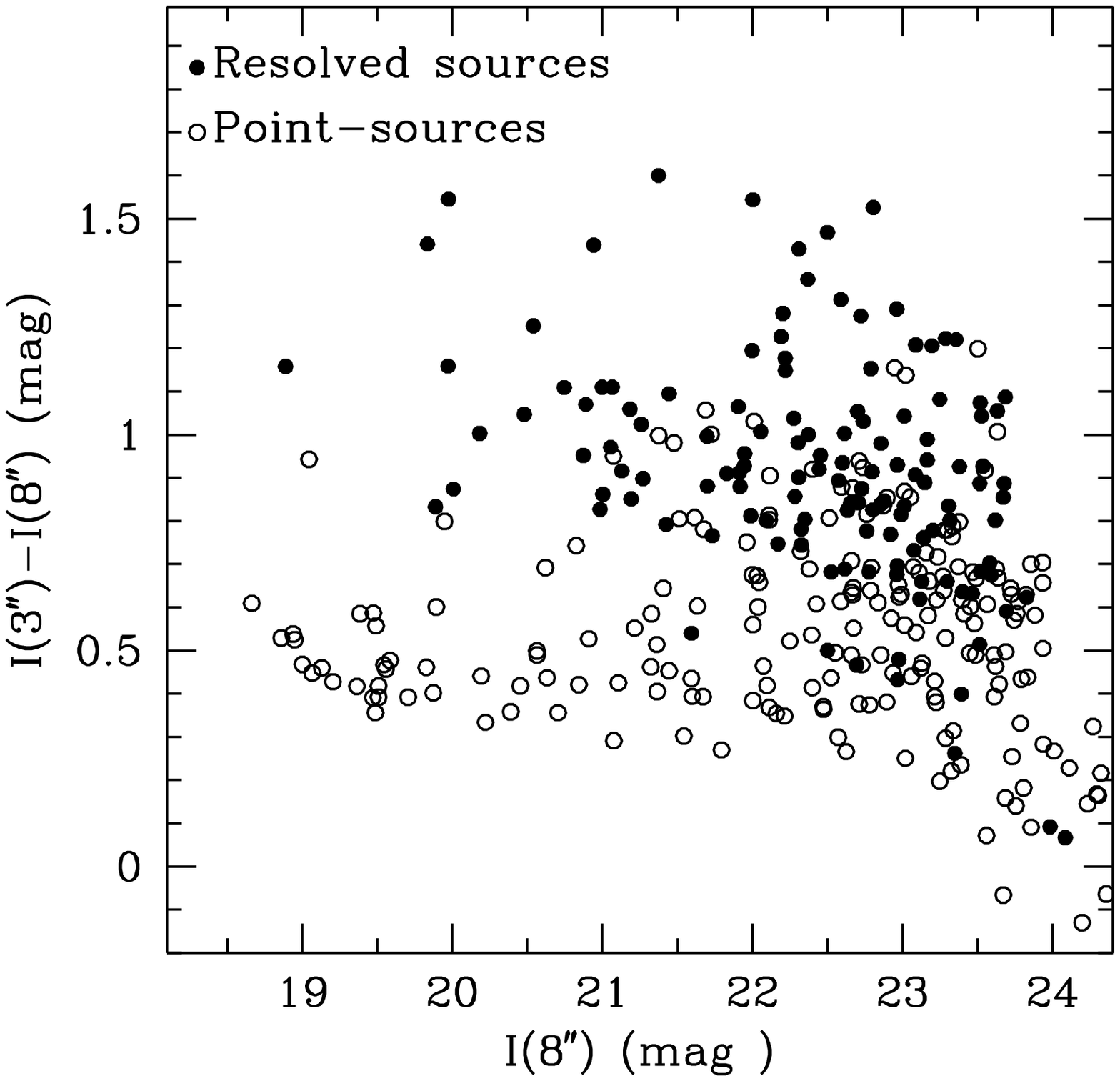,width=0.5\textwidth,angle=0}} 
 \caption
{Size parameter against $I$-band magnitude. The size parameter,
estimating the extent of a source, is defined as the $I$-band magnitude
difference within 3 and 8\,arcsec radius apertures. It is clear that
visually classified resolved sources (filled circles) are reasonably well 
separated from point-sources (open circles), although there is 
increasing scatter at fainter magnitudes.}\label{fig_svsm}    
\end{figure}

Globular cluster candidates are selected on the basis of their photometric
colours if they lie within the regions indicated in Figures \ref{fig_bv_bi}
and  \ref{fig_bv_vi}, adopting a $3\sigma$ uncertainty on each colour given
by the PHOT software. A total of 374 candidate GCs are selected. These
objects are then visually inspected to discard resolved sources, likely to
be galaxies,  and sources lying near hot pixels. The final list consists of 
205 candidate GCs. Although  visual inspection is not an objective method
for identifying point sources, we believe that to the first approximation
we successfully eliminate background galaxies from the GC catalogue.  This
is demonstrated in Figure \ref{fig_svsm} plotting an estimator of the
spatial extent of a source as a function of $I$-band magnitude. The size of
a source is estimated by the difference between the $I$-band magnitude
measured within 3 and 8\,arcsec radius apertures. It is clear from Figure
\ref{fig_svsm} that visually classified resolved sources are separated from
point-like objects, particularly at bright magnitudes. At fainter
magnitudes ($I>23$\,mag), however, the visual classification scheme becomes
more uncertain.  Also   note that point sources brighter than
$I\approx20$\,mag are likely to have saturated cores.    

Additionally, we apply a bright magnitude cutoff set to $\approx2$\,mag
brighter than  the most luminous  Milky Way GC ($\Omega$-Cen;
$M_{V}=-10.29$\,mag), shifted  to the distance of NGC\,6702. This is
because NGC\,6702 has a young stellar component, with an estimated
spectroscopic age  of $\approx 2$\,Gyrs (Terlevich \& Forbes
2000). Therefore, NGC\,6702 may possess a number of young metal--rich
GCs, expected to be significantly brighter than the old metal--poor
GCs.  For example, the Worthey (1994) models predict a 2.0\,Gyrs old
stellar population to be brighter by as much as 1.5\,mag in the
$I$-band   compared to an old metal--poor stellar population
(15\,Gyrs, $\mathrm[Fe/H]=-1.5$) assumed to be representative of the
Milky Way  GCs.  The bright  magnitude  cutoffs are thus set to
$B=22.5$, $V=21.5$  and $I=20.5$\,mag.  Sources brighter than this are
almost certainly stars. In the rest of the paper, unless otherwise
stated,  we consider GCs brighter than the $80\%$ completeness limit
for point sources. Therefore, the  magnitude range of the candidate 
GC sample is  $22.5<B<26.1$, $21.5<V<25.3$  and  $20.5<I<23.9$\,mag.
The colour and magnitude selected  and visually inspected  sample
comprises a total of 151 GC candidates.  It should be noted that
including  GCs brighter than the $50\%$ completeness limit for point
sources does not change any of our conclusions.  

\begin{figure} 
\centerline{\psfig{figure=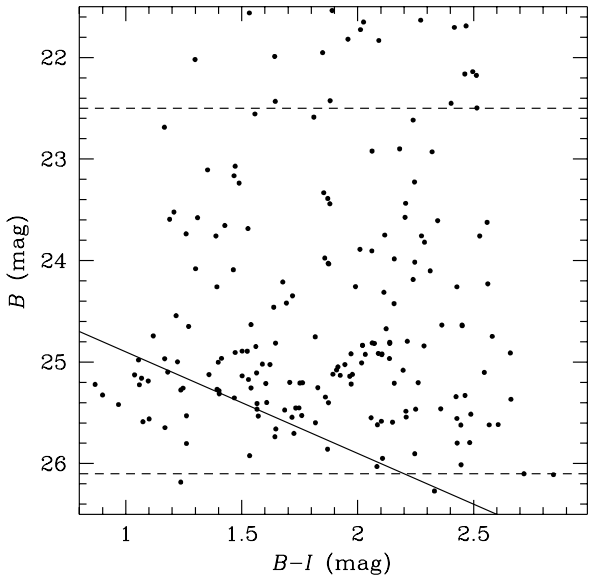,width=0.5\textwidth,angle=0}} 
 \caption
{Colour--magnitude diagram for the 205 candidate GCs, after colour
selection and visual inspection. No magnitude cutoff has been applied. 
The continuous line corresponds to the completeness 
limit $I=23.9$\,mag. The dashed
lines are the $B$-band magnitude cutoffs for the selection of the final
candidate GC sample.}\label{fig_CM}    
\end{figure}

Figure \ref{fig_CM} shows the colour-magnitude diagram of all the colour
selected and visually inspected  GC candidates before applying any
magnitude cuts. Objects  have a wide range of colours but there is some
evidence for two concentrations  at $B-I\approx1.5$  and
$B-I\approx2.3$\,mag. This will be further discussed in section
\ref{sec_res}.  

Although we attempt to minimise the fraction of foreground stars and
background galaxies in the final GC sample, it is likely that 
a number of contaminating sources are also selected. 

The number counts and colour distribution of foreground stars are estimated
using the model developed by Bahcall \& Soneira (1980)\footnote{The code
for the star count  model can be obtained from
http://ascl.net/bsgmodel.html}. In this model the Galaxy consists of both a 
disk and a spheroid component from  which the distribution  of stars in the
$B$ and $V$-bands as well as the $B-V$ colour distribution are
calculated. To predict both the star counts in the $I$-band and the $B-I$
and $V-I$ colour distributions we use the empirical tight relation between
the $B-V$ and  $V-I$  colours for stars (Bessel 1990).  We predict about
120 stars within the field--of--view in the magnitude range $22.5<B<26.0$,
$21.5<V<25.5$ and $20.5<I<24.0$\,mag. However, the colour selection
(assuming a $3\sigma$ uncertainty of 0.2\,mags) eliminates the majority of
stars from the final catalogue. Indeed, the model of Bahcall \& Soneira
(1980), after both magnitude and colour selection, predicts a total of
about $\approx20$ stars in the final GC sample (total of 151
sources). Therefore, the star contamination is about $13\%$.

\begin{figure} 
\centerline{\psfig{figure=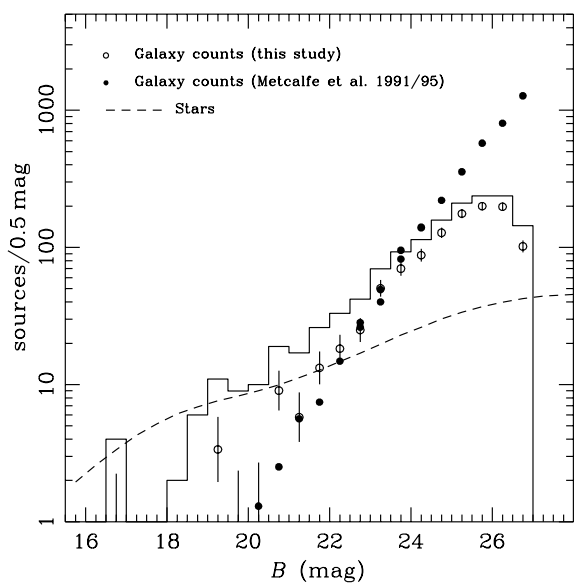,width=0.5\textwidth,angle=0}} 
 \caption {Differential number counts against $B$-band magnitude for all
the sources  detected in the NGC\,6702 $B$-band frame (histogram). The
dashed line is the star counts predicted  by the model of Bahcall \&
Soneira (1980). The open circles are the estimated number of galaxies on
the NGC\,6702 $B$-band frame (see   text). Filled circles are the $B$-band
galaxy counts of Metcalfe et al. (1991, 1995). It is clear that at bright
magnitudes the estimated number of galaxies detected on the NGC\,6702
$B$-band frame are in fair agreement with those predicted by Metcalfe et
al. (1991, 1995). At faint magnitudes  ($B>25.0$\,mag) incompleteness is
affecting the present galaxy  
sample.}\label{fig_bcounts}   
\end{figure}

\begin{figure} 
\centerline{\psfig{figure=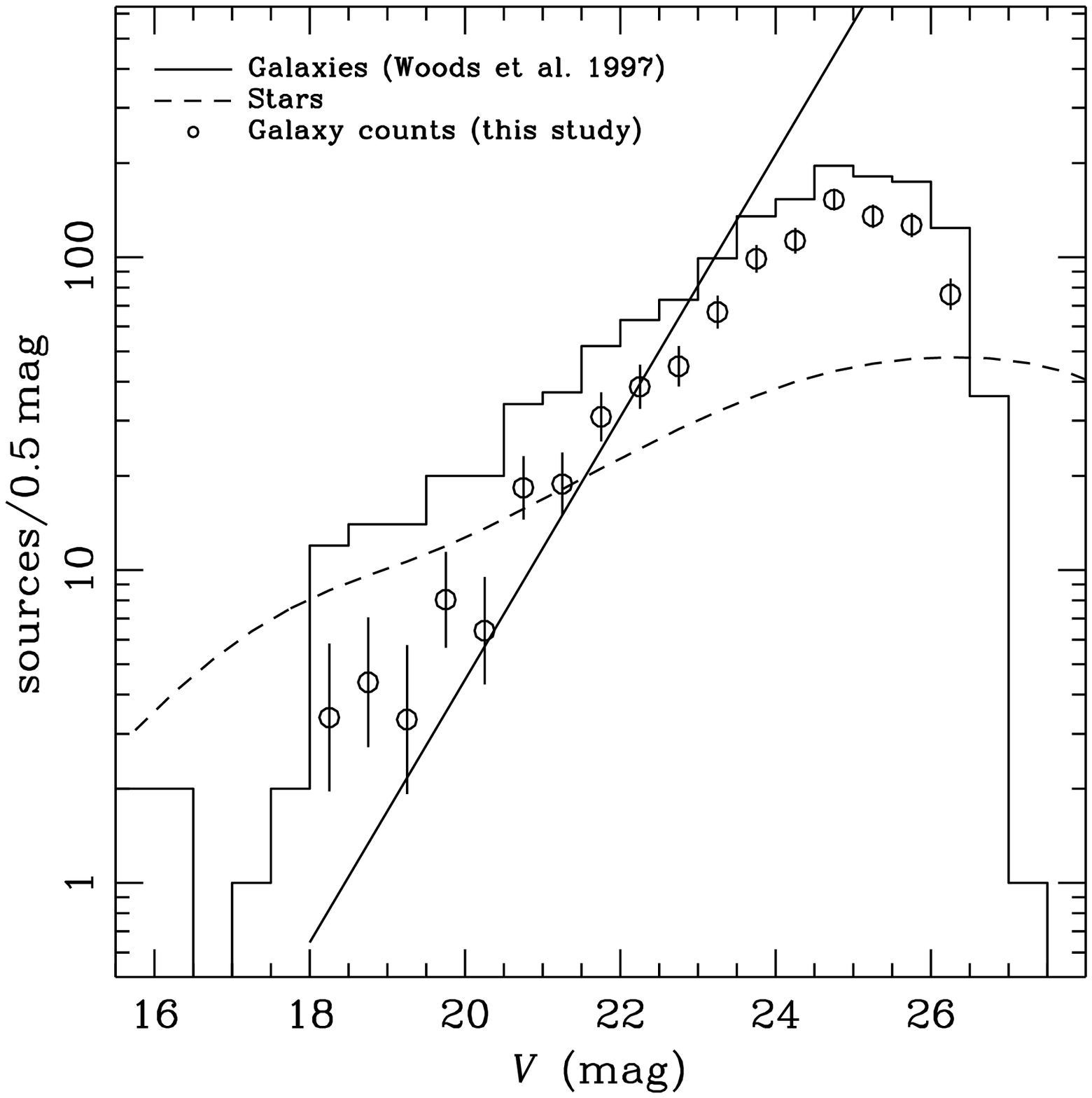,width=0.5\textwidth,angle=0}} 
 \caption
{Differential number counts against $V$-band magnitude for all the sources
detected in the  NGC\,6702 $V$-band frame (histogram). The dashed line is
the star counts predicted  by the model of Bahcall \& Soneira (1980). The
open circles  are the estimated number of galaxies on the NGC\,6702
$V$-band frame (see  text). The continuous line is the best fit to the
faint $V$-band galaxy counts estimated by Woods \& Fahlman (1997). At
bright magnitudes the estimated number of galaxies detected on the
NGC\,6702 $V$-band frame  are in fair agreement with those predicted by
Woods \& Fahlman (1997). At faint magnitudes ($V>23.0$\,mag) incompleteness
is affecting the galaxy counts from our study.}\label{fig_vcounts}     
\end{figure}

\begin{figure} 
\centerline{\psfig{figure=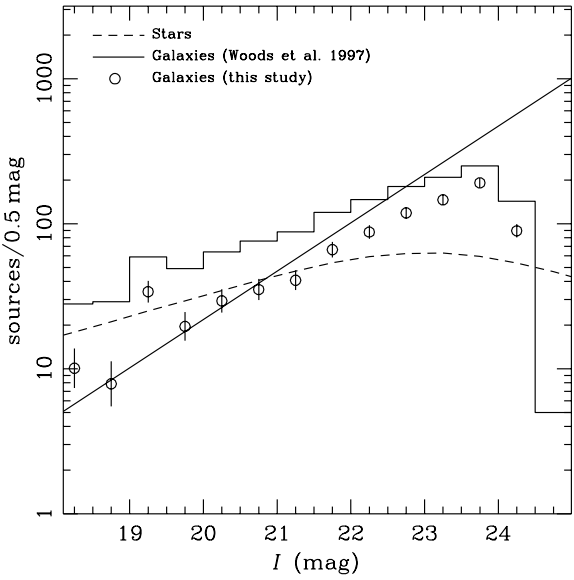,width=0.5\textwidth,angle=0}} 
 \caption
{Differential number counts against $I$-band magnitude for all the sources
detected in the NGC\,6702 $I$-band frame (histogram). The dashed line is
the star counts predicted  by the model of Bahcall \& Soneira (1980). The
open circles  are the estimated number of galaxies on the NGC\,6702
$I$-band frame (see  text).  The continuous line is the best fit to the
faint $I$-band galaxy counts estimated by Woods \& Fahlman (1997). At
bright magnitudes the estimated number of galaxies detected on the
NGC\,6702 $I$-band frame are in fair agreement with those predicted by
Woods  \& Fahlman (1997). At faint magnitudes  ($I>22.5$\,mag)
incompleteness is  affecting the galaxy counts from our
study.}\label{fig_icounts}    
\end{figure}

To quantify the contamination of the final GC sample by background galaxies 
we should first estimate the completeness limits for galaxies at different
wavebands. Since galaxies are resolved sources, their completeness limit is
expected to be at brighter magnitudes to that of point sources (e.g. GCs in
NGC\,6702). Figure \ref{fig_bcounts} plots the number counts for all the
sources detected in  our $B$-band frame (total of 1455). Also shown is the
star count prediction from the  model of Bahcall \& Soneira (1980). The
difference between the total observed source counts and the predicted  star
counts in   a given magnitude bin estimates, to the first approximation,
the number of background galaxies   detected in our field (here we ignore
the contribution from GCs to the counts). Also shown in Figure
\ref{fig_bcounts} are the $B$-band galaxy counts from Metcalfe et
al. (1991, 1995). It is clear that the predicted galaxy counts in our field
are in good agreement with those from Metcalfe et al. (1991, 1995) to the
magnitude $B\approx25$ and then decline due to incompleteness. We estimate
our $B$-band  galaxy catalogue to be about  $\approx50\%$ and $\approx30\%$
complete at $\approx25.5$ and $\approx26.0$\,mag  respectively. Figures
\ref{fig_vcounts} and \ref{fig_icounts} are similar to Figure
\ref{fig_bcounts} for the $V$ and $I$-bands respectively.  From these
Figures, we estimate our $V$ and $I$-band galaxy catalogues to be
$\approx50\%$ complete at $V\approx24.5$ and $I\approx23.5$\,mag
respectively.  

Ignoring GCs when calculating the number of galaxies within the
field--of--view at $B$, $V$ and $I$--bands, overestimates the galaxy numbers
at faint magnitudes. Consequently, the  above mentioned completeness limits
are  conservative estimates. 

To quantify the contamination of the GC sample due to background galaxies
we use the multiwavelength ($UBVI$) faint galaxy catalogue of Lilly, Cowie
\& Gardner (1991) complete to $I_{AB}=24.5$. The AB system is defined 
as $AB=48.6-2.5\log f_{\nu}$, where $f_{\nu}$ is the flux cgs units. The
transformations from the AB system are: 
$B\approx B_{AB}+0.17$, $V\approx V_{AB}$ and $I\approx I_{AB}-0.48$ (Lilly,
Cowie \& Gardner 1991). Using this galaxy sample we predict a
total of $\approx530$ galaxies with $22.5\lesssim B \lesssim 26$,
$21.5 \lesssim V \lesssim 25.5$ and  $20.5 \lesssim I \lesssim 24$\,mag
after converting to  our magnitude system and scaling to  
the area of the LRIS field--of--view ($\approx31$\,arcmin$^{2}$). Taking
into account the galaxy incompleteness of our sample, estimated from
Figures \ref{fig_bcounts}--\ref{fig_icounts},  we predict a total of
$\approx320$  galaxies within the above mentioned magnitude limits. Colour
selection, using the  aperture colours of Lilly, Cowie \& Gardner (1991)
and assuming a $3\sigma$ uncertainty of 0.2\,mags, reduces the number of
contaminating background galaxies in our candidate GC sample to
$\approx160$.  

However, as demonstrated in Figure \ref{fig_svsm}, visual inspection is
efficient in removing extended sources from the sample especially at bright
magnitudes. Indeed, a total of $\approx120$ resolved sources with $22.5
\lesssim B \lesssim 26$, $21.5 \lesssim V \lesssim25.5$ and $20.5 \lesssim
I \lesssim 24.0$\,mag have been identified, leaving about $\approx40$
unidentified galaxies within the final GC sample. This corresponds to
a contamination of the GC candidate catalogue (total of 151) by galaxies of 
$\lesssim25\%$. We stress that this fraction is an 
upper limit.  In particular, we simulated galaxy bulges following
$r^{1/4}$--law profile, with absolute luminosity and effective radius of
$M_{B}$=--21.0\,mag and $r_{eff}=4$\,kpc respectively, typical of  $L^{*}$
ellipticals. We found that only half  of them are detected to the limit
$B=25.5$\,mag and this fraction drops to $10\%$ at $B=26.0$\,mag. Moreover,
disk-dominated galaxies will be missed by the detection algorithm at even
brighter magnitudes. These surface brightness effects are likely to be
significantly less severe in the deep surveys used to determine the
galaxy counts at faint magnitudes. Indeed these  surveys have
effective exposure times as much as 10 times longer than those   
used here and are therefore, specifically designed to detect sources
at faint magnitudes. 
Since bulges and compact galaxies account for less than $20\%$ of the
galaxy population at faint magnitude limits (Elson \& Santiago 1996),
we anticipate the  galaxy contamination of our final candidate GC
sample to be much less than $\approx 25\%$.   

%Mv=-10.29, Mb=-9.51, Mi=-11.34

\section{Results}\label{sec_res}

\subsection{Colour distributions}

\begin{figure} 
\centerline{\psfig{figure=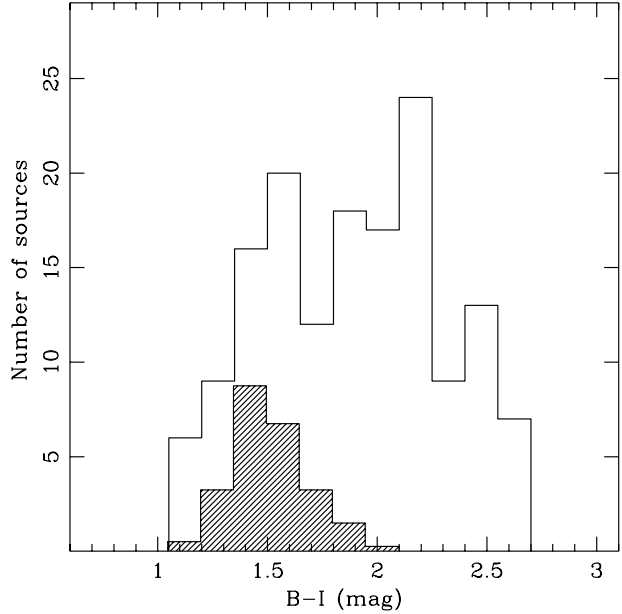,width=0.5\textwidth,angle=0}} 
 \caption
{$B-I$ colour distribution for the candidate globular clusters, 
selected on the basis of their colours (see Figures \ref{fig_bv_bi} and
\ref{fig_bv_vi}) and magnitudes. Also shown is the colour distribution of
Galactic GCs (hatched histograms). The blue peak of the NGC\,6702 GCs  lies
close to the Milky Way GC distribution. }\label{fig_hist_BI}         
\end{figure}

\begin{figure} 
\centerline{\psfig{figure=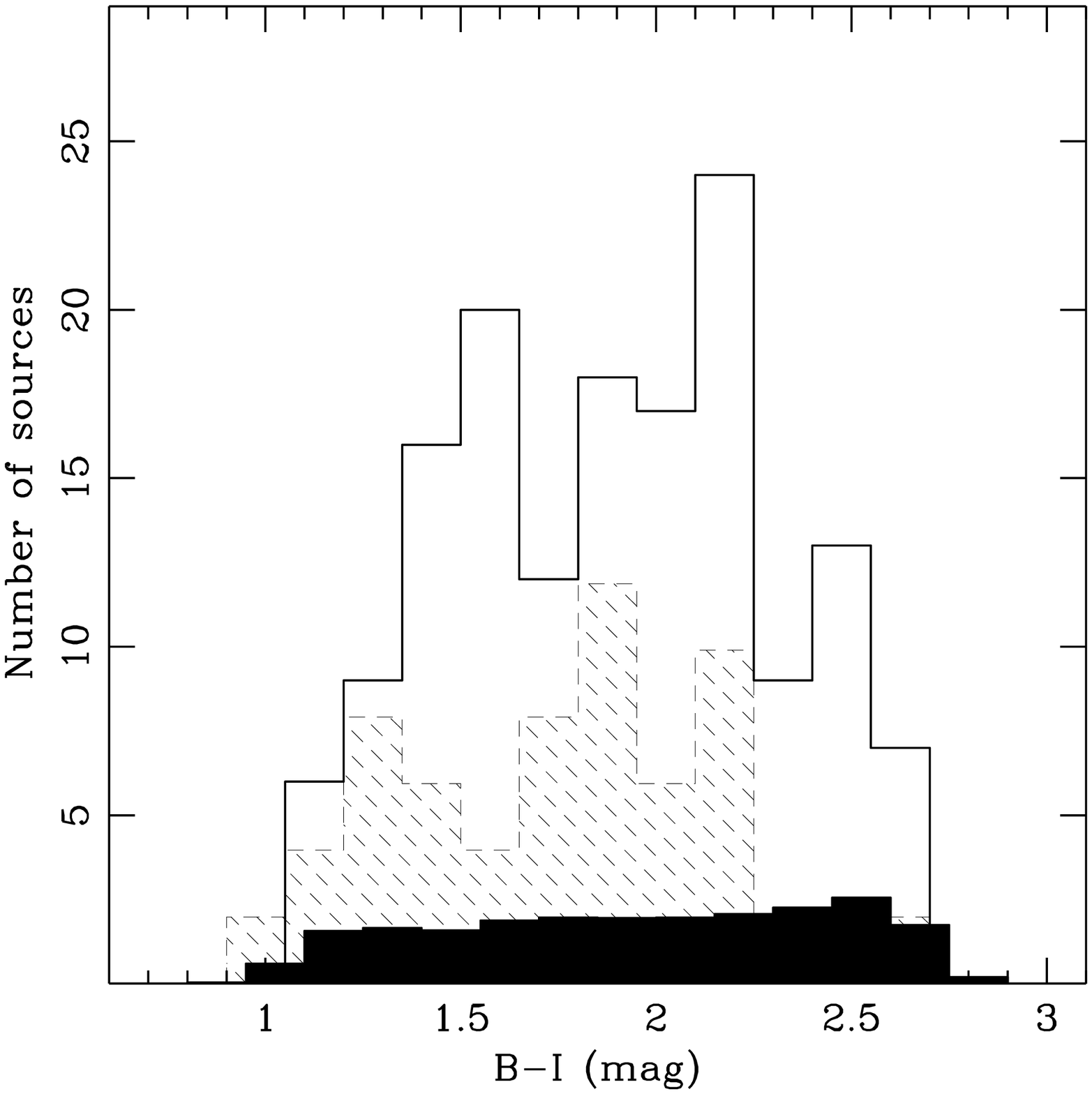,width=0.5\textwidth,angle=0}} 
 \caption
{$B-I$ histogram for the candidate globular clusters. The 
filled histogram is the distribution of contaminating stars 
predicted from the models of Bahcall \& Soneira (1980). The hatched
histogram is the predicted colour distribution of the contaminating
galaxies within the candidate GC sample.  
}\label{fig_hist_BI1} 
\end{figure}

The $B-I$ colour distribution for the final GC sample is plotted in
Figure \ref{fig_hist_BI}. There is evidence for bimodality with the peaks
of the distribution at $B-I\approx1.50$ and  $B-I\approx2.15$. Indeed the KMM 
statistical test (Ashman et al. 1994) rejects the single Gaussian model at
a confidence level better than $99.99\%$, giving peaks at $B-I=1.52$ and
2.16 respectively. It should be noted  that this result is
obtained after clipping objects with $B-I>2.6$\,mag, likely to be
background galaxies (3 out of 151; see Figure \ref{fig_hist_BI1}). However,
this is not expected to  artificially  increase the likelihood of the two
Gaussian fit. In particular, Ashman et al. (1994) argue that clipping of
the data before applying the KMM test is sometimes desirable to remove
background contamination.  
Also shown in  Figure \ref{fig_hist_BI} is the reddening--corrected colour 
distribution of  Galactic GCs  that comprise  old metal--poor GCs.  
The blue peak of NGC\,6702 GCs lies close to the  Milky Way GC
distribution.      

Figure \ref{fig_hist_BI1} compares the candidate GC $B-I$ colour
distribution with the colours of contaminating stars and galaxies (after
being selected on the basis of their magnitudes and colours) predicted 
in the previous section. It is clear that
foreground stars do not significantly modify the observed candidate GC
colour distribution. Background galaxies, however, have a larger effect on  
the observed colour distribution, having colours similar to those of the
red GC peak. 
Nevertheless, the fraction of contaminating galaxies within the sample is
expected to be sufficiently small that it  will not significantly modify
the position of the peaks of the GC colour distribution.  We further
investigate this by subtracting  the predicted galaxy colour distribution
from that of the GC candidates in Figure \ref{fig_hist_BI1}. The KMM
statistical test is then applied to the resulting distribution. A single
Gaussian model is rejected at a confidence level better than $99.99\%$,
while the peaks of the bimodal distribution are at $B-I=1.56$ and
2.27\,mag respectively. Although both peaks are redder than those 
recovered using the full candidate GC colour distribution, the difference
is sufficiently small that does not alter any of our conclusions.

\begin{figure} 
\centerline{\psfig{figure=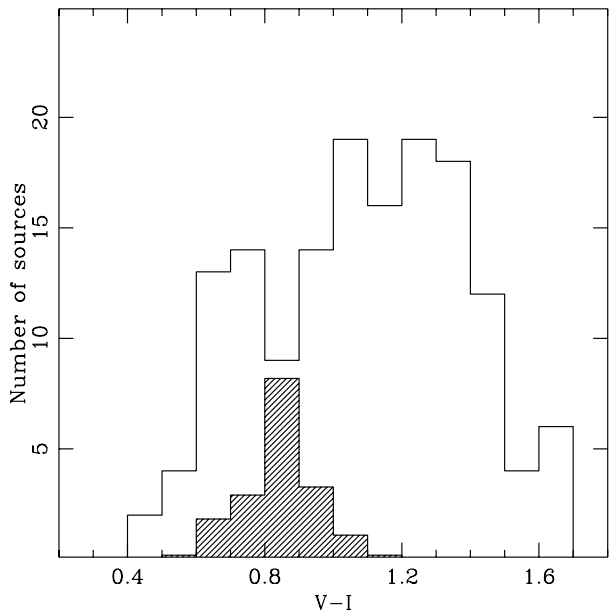,width=0.5\textwidth,angle=0}} 
 \caption
{$V-I$ colour distribution for the candidate globular clusters, 
selected on the basis of their colours and magnitudes. Also shown are the
colour distributions of Galactic GCs (hatched
histograms). }\label{fig_hist_VI}          
\end{figure}

\begin{figure} 
\centerline{\psfig{figure=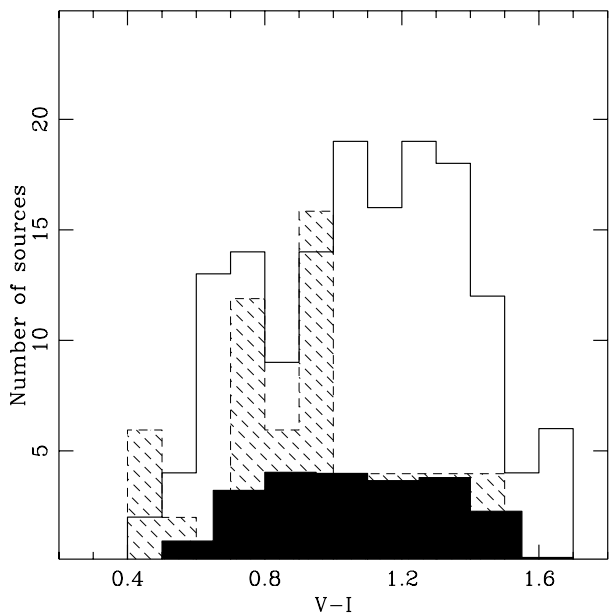,width=0.5\textwidth,angle=0}} 
 \caption
{$V-I$ histogram for the candidate globular clusters. The 
filled histogram is the distribution of contaminating stars 
predicted from the models of Bahcall \& Soneira (1980). The hatched
histogram is the predicted colour distribution of the contaminating
galaxies within the candidate GC sample.   
}\label{fig_hist_VI1} 
\end{figure}

Figure \ref{fig_hist_VI} plots the $V-I$ colour histogram of both the
NGC\,6702 GC candidates and the Galactic GCs. There is no strong evidence 
for bimodality, with the KMM test finding no improvement of the two-group
fit over a single Gaussian. This seems surprising since the $B-I$ colour
distribution shows clear evidence for bimodality. 
It is likely that the absence of bimodality in Figure \ref{fig_hist_VI} is
due to the smaller wavelength baseline of $V-I$ compared to the
$B-I$ colour, providing less metallicity leverage. The expected 
difference between the peaks in $V-I$ is smaller (i.e. $\Delta
(V-I)\approx0.3$\,mag; see next section) than that in the $B-I$ histogram
and therefore  random photometric errors can more easily smear out any 
bimodality.  Contamination of the GC candidate sample by stars and galaxies
may also affect the significance of the detection of bimodality in the
$V-I$ colour distribution. This is demonstrated in Figure
\ref{fig_hist_VI1}, plotting the $V-I$ colour distribution of
contaminating stars and galaxies (after being selected on the basis of
their magnitudes and colours).

\subsection{Age of the NGC\,6702 Globular Clusters}\label{sec_age}

To constrain the age and metallicity of NGC\,6702 GC system we employ a method
similar to that developed by Whitmore et al. (1997) that uses the
colour-magnitude relation of GCs. The basic assumptions of the method are  
discussed in detail by Whitmore et al. (1997). In brief, it is assumed
that the GCs comprise two distinct sub-populations of different age and
metallicity but with the same initial mass functions. An additional
assumption  is that both sub-populations have similar numbers of
GCs. The older population is taken to have properties  similar to
those of the Milky Way GCs and therefore consists of old (15\,Gyrs)
metal--poor ($\mathrm{[Fe/H]=-1.5}$) GCs. The younger population is
formed at a later stage of the evolution of the galaxy, through either
a merger event (e.g. Ashman \& Zepf 1992) or a secondary {\it in situ}
star-formation burst  (Forbes et al. 1997). The young GCs, assumed
to be metal--rich, are initially bluer than the old metal--poor ones
but become redder after about 1--2\,Gyrs. Similarly, the newly formed
GCs are much brighter than the pre-existing metal--poor population but
become fainter at later stages as they evolve passively. Therefore,
the difference between the colours and magnitudes of the two GC
populations can be used to constrain the age of the new GCs.  

\begin{figure} 
\centerline{\psfig{figure=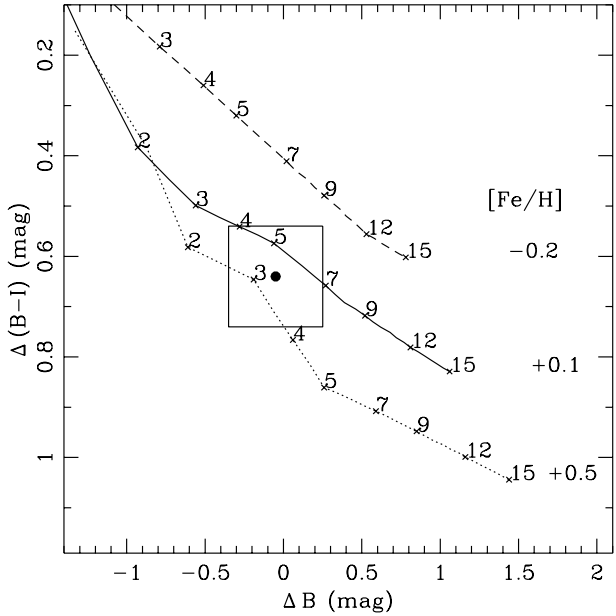,width=0.5\textwidth,angle=0}} 
 \caption
{$\Delta(B-I)$ against $\Delta B$ diagram for the Worthey (1994)
models. Each curve corresponds to a different metallicity marked on the
right of the curve. The crosses and the numbers on the curves indicate
different ages in Gyrs. $\Delta(B-I)$ is defined as the colour difference
between the red and the blue old (15\,Gyrs) metal--poor
($\mathrm{[Fe/H]=-1.5}$) GC populations. $\Delta B$ is defined as the
magnitude difference of the 10th brightest globular cluster of the red and 
blue populations. The dot corresponds to the  NGC\,6702 GCs
($\Delta(B-I)=0.64\pm0.10$, $\Delta B=-0.05\pm0.29$), while the rectangle
shows the $1\sigma$ uncertainty region.}\label{fig_worthy_b} 
\end{figure}

\begin{figure} 
\centerline{\psfig{figure=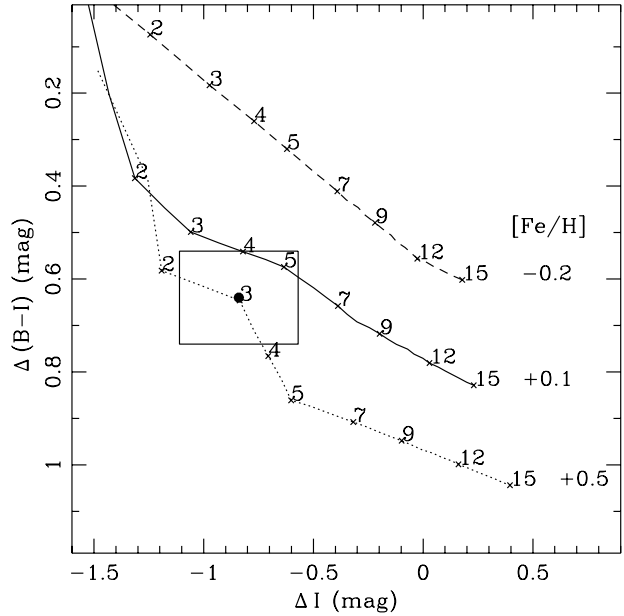,width=0.5\textwidth,angle=0}} 
 \caption
{$\Delta(B-I)$ against $\Delta I$ diagram for the Worthey (1994)
models. Each curve corresponds to a different metallicity marked on the
right of the curve. The crosses and the numbers on the curves indicate
different ages in Gyrs. $\Delta(B-I)$ is defined as the colour difference
between the red and the blue old (15\,Gyrs) metal--poor
($\mathrm{[Fe/H]=-1.5}$) GC populations. $\Delta I$ is defined as the
magnitude difference of the 10th brightest globular cluster of the red and 
blue populations. The dot corresponds to the NGC\,6702 GCs
($\Delta(B-I)=0.64\pm0.10$, $\Delta I=-0.84\pm0.27$) while the rectangle
shows the $1\sigma$ uncertainty region.}\label{fig_worthy_i}            
\end{figure}

Here we employ the models of Worthey (1994) to predict the colour and
magnitude difference of a young GC population relative to an old 
(15\,Gyrs) metal--poor ($\mathrm{[Fe/H]=-1.5}$) population.  The predicted
colours of this old stellar  population are $V-I=0.921$ and $B-I=1.564$.  
The model predictions for the $B-I$ colour against both $B$ and $I$
magnitudes are shown in Figures \ref{fig_worthy_b} and  \ref{fig_worthy_i}
respectively. For NGC\,6702 the difference $\Delta(B-I)$ between the
colours of the two GC populations is taken as the difference between
the peaks of the bimodal $B-I$ distribution  in Figure
\ref{fig_hist_BI}. The blue peak in that figure is assumed to
correspond to the  old metal--poor population. This is a reasonable
assumption since the blue peak overlaps with the Milky Way GCs fairly
well.  The peak colour difference from the KMM statistical test is
$\Delta(B-I)=0.64\pm0.10$,  where the error is estimated assuming an 
uncertainty of 0.07\,mag in defining the peaks of the bimodal
distribution. 

The $B$ and $I$-band  magnitude difference, $\Delta B$ and $\Delta I$,
between the two GC populations is estimated following the method of
Whitmore et al. (1997) by taking the offset between the 10th brightest
cluster of the blue  and red  subpopulations. The blue and red
subpopulations are defined to have colours in the range $1\le B-I<
1.8$ and $1.8\le B-I< 3$ respectively.  The uncertainties in $\Delta
B$ and $\Delta I$ are estimating using the bootstrap resampling
technique. We find  $\Delta B=-0.05\pm0.29$ and $\Delta
I=-0.84\pm0.27$, i.e. the young GCs are brighter than the old ones.   

The results are shown in Figures \ref{fig_worthy_b} and
\ref{fig_worthy_i}, and suggest that NGC\,6702 GCs have solar to supersolar
metallicity  and an age in the range 2-5 Gyrs. Also we note that despite
the errorbars, in both figures the data point lies close to the age of
3\,Gyrs and [Fe/H]=0.5 model prediction. Moreover, although there are
uncertainties in the calculation of the age of the newly formed GCs, our
estimate is in fair agreement with the galaxy's spectroscopically estimated
age and metallicity of $\approx2$\,Gyrs and
$\mathrm{[Fe/H]\approx0.5}$ respectively (Terlevich \& Forbes
2000). 

To assess the sensitivity of the age and metallicity estimates to
the adopted single stellar population model of Worthey, we repeat our
calculations using the models developed by Bruzual \& Charlot (1996)
and Vazdekis et al. (1996) with a Salpeter IMF. Both models predict an
age of about 2\,Gyrs for the recently formed GC population. The
predicted metallicity of $\mathrm{[Fe/H]>0.5}$ is higher than that
from the Worthey model. Despite, the discrepancies, it is clear that
all three  models used here predict that the blue GCs sub-population
is relatively young with a supersolar metallicity.

Additionally, assuming an  age of 3\,Gyrs and [Fe/H]=0.5 for
the red GCs, the Worthey (1994) models predict a $V-I$ colour
difference between the two GC subpopulations of $\Delta
(V-I)\approx0.3$\,mag. As discussed in the previous section such a 
small colour difference is difficult to detect with the present dataset. 

\subsection{The surface density and specific frequency}\label{sec_SN}

\begin{figure*} 
\centerline{\psfig{figure=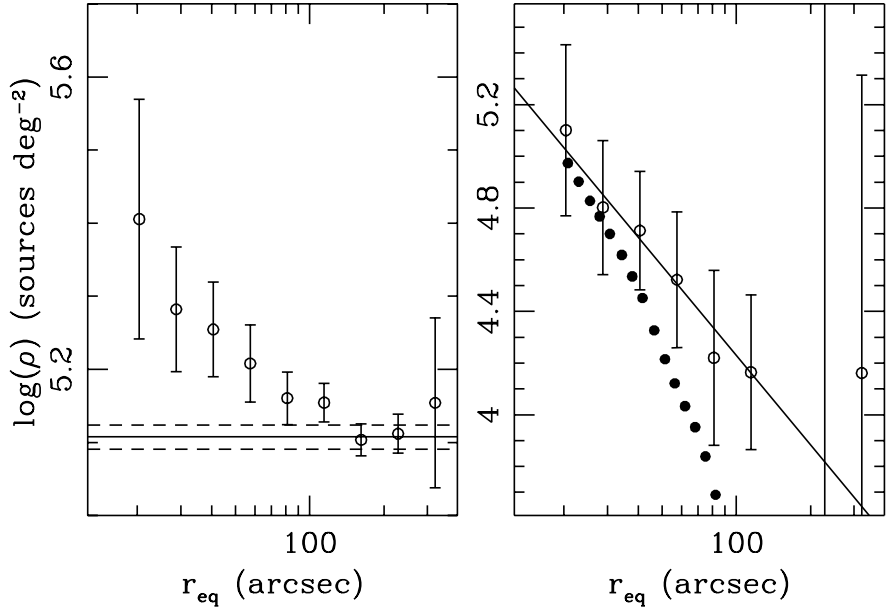,width=0.7\textwidth,angle=0}} 
 \caption
{{\bf (a)} Left panel: the surface density profile of all the sources
 detected in the NGC\,6702 $B$-band frame. The continuous lines is the
 mean background surface density, estimated by combining the counts in the
 bins at galactocentric distances $>120^{\prime\prime}$. The dashed lines
 are the  errors around the mean assuming Poisson statistics. {\bf (b)}
 Right panel:  the surface density  profile of the globular cluster
 candidates (open  circles), derived by statistically subtracting the
 background surface  density level from the density profile of the left
 panel. The best fit  power law to the observed profile (continuous line)
 has the form  $\rho\propto r^{-1.15\pm0.52}$. The filled circles are the
 $B$-band  surface brightness profile of NGC\,6702 in arbitrary units. The
 starlight  is more concentrated  than the GC system.}\label{fig_den_prof}    
\end{figure*}

We study the surface density profile, $\rho(r)$, of the NGC\,6702 GCs using 
all  the sources detected in the $B$-band to  $B=26.1$\,mag (total of 
1237), the $80\%$ completeness  limit for point sources. 
We use the $B$-band detections instead of the colour selected GC sample
because the selection criteria (i.e. completeness limits) are  easier to
quantify and do not depend on the GC colours.  
The sample is binned into elliptical rings centred on NGC\,6702 and having
the same mean ellipticity and position angle as the galaxy ($\epsilon=0.2$,
PA=$145^{\circ}$).  The counts in each bin are corrected for geometric
incompleteness  due to foreground saturated stars, bad columns and the
limited size of the field--of--view. Additionally,  we mask out the central
saturated part of NGC\,6702 for radii $r_{eq}<21^{\prime\prime}$.  
The correction for magnitude incompleteness to the counts
in individual bins was  found to be small ($\approx3\%$) and was ignored.
The density profile is plotted in Figure \ref{fig_den_prof}(a).  
The background surface density level, $\rho_{bg}$, was determined by
combining the counts within the outermost three bins lying at
$r_{eq}>120^{\prime\prime}$ (corresponding to  galactocentric distances
$>37$\,kpc) where the density profile reaches a constant level. We find
$\rho_{bg}=(1.28\pm0.05)\times10^5\mathrm{\,sources\,deg^{-2}}$ (assuming
Poisson statistics) or
$1114\pm44$ sources over the LRIS field--of--view of  
31.2\,arcmin$^{2}$ (after applying geometrical corrections). The total
number of objects to $B=26.1$\,mag is 1237 and therefore the number of
candidate GC is estimated to  be 123$\pm$56 (Poisson statistics). In section
\ref{sec_sel} we found a total of 151 candidate GCs, selected on the basis of
their colours and magnitudes in the range $22.5<B<26.1$\,mag. This is in
fair agreement with the number of candidate GCs estimated here. 

Figure \ref{fig_den_prof}(b) shows the candidate GC density profile,
estimated  by statistically subtracting the constant background surface
density level from the density profile of all the sources detected on the 
$B$-band frame to $B=26.1$\,mag. Assuming a power law density profile of
the form  $\rho\propto r^{\alpha}$ for the candidate GCs, we find 
$\alpha=-1.15\pm0.52$. This slope for the GC system of NGC\,6702 is
consistent with that for other galaxies of a similar magnitude (Forbes
et al. 1997).  Also shown in Figure \ref{fig_den_prof}(b) is the
$B$-band  surface brightness profile of NGC\,6702. It is clear that the
starlight of NGC\,6702 drops more steeply than the GC density
profile as is often seen in other  galaxies (Harris 1991; Durrell
et al. 1996). This is often attributed to further dissipative collapse of
the host galaxy after the older GCs had formed and before the halo stars
formed (e.g. Harris 1986). The radial distributions of the red and blue
subpopulations of the GC candidates selected on the basis of their colours
and magnitudes have also been investigated and have been found to be
consistent within the uncertainties.  

The globular cluster specific frequency $S_{N}$ is defined (Harris 1991)
\begin{equation}
S_N=N_T\times10^{0.4*(M_V+15)},
\end{equation}
where $N_T$ is the total number of globular clusters in the galaxy and
$M_V$ is the host galaxy absolute  $V$-band luminosity. To calculate $N_T$
we assume that the  $B$-band luminosity function of NGC\,6702 GCs is a
Gaussian with the same parameters as that of the Milky Way GCs
($M_{B,peak}=-6.50$\,mag and $\sigma_B=1.13$). We note that differences in
the peak $B$-band magnitude of the blue and red GC subpopulation luminosity
function are small (see section \ref{sec_age}). At the distance of
NGC\,6702  $M_{B,peak}$ corresponds to $B$=27.5\,mag. Previously, we found
that the number of GCs to the limiting magnitude $B$=26\,mag is
123$\pm$56. Therefore integrating over the entire area of the luminosity
function we find $N_T=1776\pm926$. This implies $S_N=2.3\pm1.1$, assuming
$M_V=-21.9$\,mag for NGC\,6702.   

It is an interesting exercise to predict the evolution of $S_N$ with time
to investigate whether NGC\,6702 will resemble normal ellipticals within
a few Gyrs.  For this calculation we assume that the total number of GCs is
conserved and that the change in $S_N$ is due to passive evolution of the
host galaxy. Furthermore, we assume that during the most recent major
star-formation event, that took place about 2\,Gyrs ago, $10\%$ of the
NGC\,6702 mass (see section \ref{sec_disc} below) has been converted into
metal--rich ([Fe/H]=+0.5) stars. After 10\,Gyrs this new stellar population
will fade in the $V$-band by $\Delta V=0.18$\,mag (Worthey 1994),
corresponding to $S_{N}=2.7$. Harris, Harris \& McLaughlin (1998)  found a
mean of $S_{N}=5.4\pm0.3$ for large ellipticals in the Virgo, Fornax and
the NGC\,5846 group (excluding cD galaxies).  However, exclusion of
Bright Cluster Galaxies (BCGs)  results in a mean $S_{N}=3.3\pm0.2$, in
better agreement, although somewhat larger, with the estimated
$S_{N}\approx2.7$ for NGC\,6702 after 10\,Gyrs. However, the estimated
$S_{N}$ evolution is sensitive to the galaxy mass fraction involved in the
most recent major star-formation event, which in turn is uncertain and
difficult to constraint. We calculate an upper limit for the NGC\,6702
$S_{N}$ after 10\,Gyrs by assuming the extreme scenario  where as much as
$50\%$ of the galaxy mass has been turned into stars.  We estimate a fading
of $\Delta V=0.82$\,mag (Worthey 1994) within 10\,Gyrs, corresponding to an
upper limit of $S_{N}=4.9$. Therefore, we conclude that the NGC\,6702
$S_{N}$, although lower than that of bright cluster ellipticals, it is
likely to be similar to that of `normal' (old) present-day ellipticals
(excluding BCGs) within the next 10\,Gyrs.  The calculations above are
also consistent with suggestions that galaxy interactions and mergers
although unable to account for large $S_N$ galaxies, may explain
the low values of $S_N\approx2$ in relatively isolated ellipticals
(Elmegreen 1999).

\section{Discussion}\label{sec_disc}

\begin{table} 
\footnotesize 
\begin{center} 
\begin{tabular}{lrr}
\hline 
 & NGC\,6702 & NGC\,1700\\
Type & E3 & E4\\
M$_V$ (mag)& --21.9 & --21.5\\
Environment & Field & Group\\
$(V/\sigma)^{\ast\,\,1}$ & 0.21 & 0.38\\ 
Spectroscopic & & \\
age$^{2}$  (Gyr) & 2 & 2.3\\
$\mathrm{[Fe/H]}^{2}$ & 0.5 & $>$0.5\\
$\mathrm{[Mg/Fe]}^{2}$ & 0.12 & 0.13\\
Fundamental Plane & & \\ 
residual$^{3}$ & --0.20 & --0.37 \\
$S_{N}^4$     & $2.3\pm1.1$ & $1.4^{+1.0}_{-0.6}$ \\
\hline
\multicolumn{3}{l}{$^1$Bender et al. 1992} \\ 
\multicolumn{3}{l}{$^2$Terlevich \& Forbes 2000} \\ 
\multicolumn{3}{l}{$^3$Forbes, Ponman \& Brown 1998} \\ 
\multicolumn{3}{l}{$^4$The $S_N$ for NGC\,1700 from Brown et al. 2000} \\ 
\hline
\end{tabular}
\end{center} 
\caption{Comparison between the properties of NGC\,6702 and NGC\,1700
($H_0=75\,\mathrm{km\,s^{-1}\,Mpc^{-1}}$).}\label{tab_2}  
\normalsize  
\end{table}

A comparison between the properties of NGC\,6702 and the group
proto-elliptical NGC\,1700 (Brown et al. 2000; Whitmore et al. 1997),
is shown in Table 2.  It is clear that there are quite a few
similarities between the two galaxies. Both of them have similar
optical luminosities and deviate from the Fundamental Plane in the
sense of  having younger ages on average (Forbes, Ponman \& Brown
1998). This is confirmed by their central spectroscopic ages of
$\sim$2\,Gyrs (Terlevich \& Forbes 2000). The central stars appear to
be of high metallicity ([Fe/H] $\sim$0.5) with supersolar [Mg/Fe]
abundance. Supersolar abundances have been the source of much debate
in the literature but probably indicate either a rapid star formation
time-scale or an IMF skewed towards massive stars.  Both galaxies have
radial velocity to velocity dispersion ratios, $(V/\sigma)^{\ast}$,
indicating that they  are anisotropic rotators. Bender et al. (1992)
have suggested that the last major merger for anisotropic 
galaxies involved relatively little gas, i.e. the progenitors were largely
stellar.  In the case of NGC\,6702 and NGC\,1700 there must have been
sufficient gas to support the central starburst and associated cluster
formation. The similarities above also suggest a similar formation
mechanism for these galaxies.

Moreover, our analysis above indicates the presence of two GC populations in
NGC\,6702 with  different metallicities and ages. 
It is an interesting exercise to estimate the formation efficiency of the
young (red; $1.8\leq B-I\leq 3$) GC subpopulation, assuming they formed
during the most recent star-formation burst, and compare it to that of
stars formed in the same burst.      

A robust estimator of  the GC formation efficiency is the ratio of the
total GC mass to the total baryonic galaxy mass. For this calculation we 
assume a typical GC mass of $3\times10^{5}\,M_{\odot}$ (Harris, Harris \&
McLaughlin 1998) and a mass--to--light ratio for NGC\,6702
$(M/L)_{V}=8\,M_{\odot}/L_{\odot}$, typical to that of ellipticals (Faber
\& Gallagher 1979; Binney \& Tremaine 1987). We have identified a
total of 88 red GCs ($1.8\leq B-I\leq 3.0$) to the limit
$B\approx26.0$\,mag, assumed to be  the newly formed metal--rich GCs of
NGC\,6702. We adopt a Gaussian luminosity function for this GC
subpopulation with parameters similar to those used in section \ref{sec_SN}
($M_{B,peak}=-6.50$\,mag and $\sigma_B=1.13$). This is a reasonable
assumption, since the blue and red GC subpopulations of NGC\,6702 in
section \ref{sec_age} have a similar $B$-band magnitude distribution within
few tenths of a magnitude. Integrating  over the entire  luminosity
function, assuming a completeness limit for our  sample
$B\approx26.0$\,mag, we estimate a total of 970 red GCs. Thus, the total
mass of the red GCs is
$M_{GC}=970\,\times\,3\times10^{5}\,M_{\odot}=2.9\times10^{8}\,M_{\odot}$. 
The total baryonic mass in stars of NGC\,6702, estimated using its
$V$-band luminosity and the assumed mass--to--light ratio, is found to 
be $M_{G}=3.6\times10^{11}\,M_{\odot}$. Therefore, we  estimate that
$0.08\%$ of the NGC\,6702 mass  has been turned into GCs during the  most
recent star-formation event, which is lower compared to  recent estimates
for the GC formation efficiency of 0.25\% (Harris, Harris \& McLaughlin
1998; McLaughlin 1999). It should be noted that this calculation assumes
that the estimated baryonic mass in stars represents the total mass of the
system, i.e. most of initial gas has been used for star-formation.  

Elmegreen et al. (1999), argue that GCs represent about $1\%$ of the total 
galaxy star-formation. Therefore, for the GC formation efficiency of
$0.08\%$ we expect that $\approx8\%$ of the total  NGC\,6702 mass has
turned into  field stars (not in bound clusters) during the most recent
star-formation event. Is our estimate of the mass in new stars (i.e. 8\% of
the total) consistent with the spectral line indices in the central r$_e$/8
region  (Gonzalez et al. 1993)? If we assume that the new stars formed with
a metallicity of  [Fe/H]=0.5, and that the old stars (representing 92\% of 
the central  stellar mass) have solar metallicity and an age of 15\,Gyrs,
then the new  stars would have to be slightly younger than 1\,Gyr
old. Assuming a time since the starburst of $\approx2$\,Gyrs and  applying
that  to the new  stars, their mass fraction would have to be increased to 
$\approx15\%$. Furthermore, if the  galaxy reveals evidence for 
a young stellar population outside of the  central region this will also
point towards a large mass fraction (see also Trager 1997). Clearly there
are large uncertainties involved in the above arguments, but there is some
agreement between the mass fraction of young stars estimated independently
using spectroscopic line indices and the properties of the NGC\,6702 GC
system. Moreover, it seems clear that a large fraction of new stars  formed
in NGC\,6702 a few Gyrs ago. In the merger context, it implies a  high gas
fraction for the progenitors and hence a highly dissipational event. This
is contrary to the claims of Bender et al. (1992) that anisotropic
galaxies (such as NGC\,6702 and NGC\,1700) formed by largely   {\it
dissipationless} mergers. 

Next, we  compare the fraction of galaxy starlight in newly formed dense
clusters relative to that of young field stars. To do that we add the
luminosities of the red GCs (total of 88;
$L_{GC,B}=4.9\times10^{7}\,L_{\odot}$) and then divide by the luminosity 
of the {\it young} stellar population of NGC\,6702. 
We note that estimating the total GC luminosity by extrapolating to the GC
luminosity function will not modify our results, since the main
contribution to the luminosity is from bright systems. Therefore, adding
the red GC luminosities is a simple approximation that is sufficient for
the purposes of the present study.  
Additionally, in our calculation we assume that $10\%$ by mass of the
NGC\,6702 stars  are young (2\,Gyrs) and metal rich ([Fe/H]=0.5), while the
remaining $90\%$ of the mass is in old (15\,Gyrs)  metal poor ([Fe/H]=--1.5)
stars. Using the mass--to--light ratios of the above stellar populations
predicted by the models of Worthey (1994) we estimate that $15\%$ of the
total NGC\,6702 $B$-band light originates in newly formed
stars, i.e. $L_{\star,B}=2.9\times10^{9}\,L_{\odot}$. Consequently, we
estimate $L_{GC,B}/L_{\star,B}\approx 0.02$ , i.e. $2\%$ of the young star  
$B$-band luminosity originates in bound clusters formed during the most
recent star-formation event. This is much lower than the estimates in
recent  mergers which have shown that as much as $20\%$ of the blue
galaxy light originates from GCs (Zepf et al. 1999; Meurer
1995). However, the main contribution to the  integrated starlight in
bound clusters in these systems is from young luminous clusters, the
nature of which is under debate. In particular, it is still unclear
whether these  objects are proto-globular clusters that will evolve
into {\it bona-fide} GCs (Brodie et al. 1998) or are luminous open
clusters that will be destroyed  by dynamical evolution (van den Bergh
1995). Indeed, in older merger remants, like NGC\,7252, the fraction
of light in bound clusters is somewhat lower than $20\%$ (Zepf et
al. 1999). Moreover, van den Bergh  (1994) noted that GCs account for
about $2\%$ of the total halo luminosity of both the Milky Way and the
LMC, in good agreement with our estimate for NGC\,6702.   

Clearly, the young elliptical galaxy NGC\,6702 (and also NGC\,1700) is at
an interesting evolutionary stage for further study of its GC
system. Direct measurement of individual GC metallicities with 8-10m class
telescopes would confirm (or refute) the inferred supersolar metallicities
for the new GCs and hence their age estimates. We note that the magnitude
differences $\Delta B$ and $\Delta I$ in section \ref{sec_age}, used to
constrain the GC age and metallicity, are hard to define and would benefit
from deeper imaging. Deeper imaging studies would also help determine the
GC luminosity function (e.g. Gaussian or power law) to explore possible
destruction mechanisms and would allow a detailed study of the radial
distribution of the GC subpopulations.   

\section{Conclusions}\label{sec_conc}

We have carried out $B$, $V$ and  $I$-band photometric observations of
the elliptical galaxy NGC\,6702 to explore its GC system. The
GC candidates are selected on the basis of their colour, magnitude and
morphology. We find strong evidence that the $B-I$ colour distribution
of GC candidates is bimodal with the blue peak having a colour similar
to that of the  old metal--poor Galactic GCs. Assuming that the blue GC
subpopulation is indeed old and metal--poor, we estimate that the red
GCs have an age of 2--5\,Gyrs and super-solar metallicity. Despite the
large uncertainties, this is in reasonable agreement with the
spectroscopically estimated age and metallcity of about 2\,Gyrs and
0.5 respectively for the galaxy. Thus the red GCs in
NGC\,6702 are consistent with having formed during the most recent
starburst, likely to have been triggered by a merger or an accretion
event, a few Gyrs earlier. Additionally, we find a  specific frequency for
the NGC\,6702 GC system of $S_N=2.3\pm1.1$. We predict that passive
evolution  of the NGC\,6702 stellar population formed during the most
recent star-formation event will increase the NGC\,6702 specific frequency
to $\approx2.7$ within 10\,Gyrs, in reasonable agreement with the  
mean for present-day ellipticals, excluding the largest cluster
ellipticals. Finally, using the NGC\,6702 GC properties and making
reasonable assumptions about the GC formation efficiency, we 
estimate the mass fraction of newly formed stars during the most recent
star-formation event. We find reasonable agreement with the mass fraction
estimated using stellar population synthesis models to interpret the line
indices of the underlying galaxy starlight.   

\section{Acknowledgements}\label{sec_ack}

We thank Soeren Larsen and Richard Brown for useful comments and
suggestions. Part of this research was funded by NATO Collaborative
Research grant CRG 971552.  The data presented herein were obtained at the
W.M. Keck Observatory, which is operated as a scientific partnership among
the California Institute of Technology, the University of California and
the National Aeronautics and Space Administration.  The Observatory was
made possible by the generous financial support of the W.M. Keck
Foundation. This research has made use of the NASA/IPAC Extragalactic
Database  (NED), which is operated by the Jet Propulsion Laboratory,
Caltech, under contract with the National Aeronautics and Space
Administration.


\begin{thebibliography}{} 

{\bibitem{1} Ashman K. A., Bird C. M., Zepf S. E.,
1994, AJ, 108, 2348}

{\bibitem{2} Ashman K. M., Conti A., Zepf S. E., 1995, AJ, 110, 1164}

{\bibitem{3} Ashman K. M., Zepf S. E., 1992, ApJ, 384, 50}

{\bibitem{4} Bahcall J. N., Soneira R. M., 1980, ApJS, 44, 73}

{\bibitem{5} Barnes J. E., 1992, ApJ, 393, 484}

{\bibitem{6} Bender R., Burstein D., Faber S. M., 1992, ApJ, 399, 462}

{\bibitem{7} Bertin E., Arnouts S., 1996, A\&AS, 117, 393}

{\bibitem{8} Bessel 1990, A\&AS, 83, 357}

{\bibitem{9} Binney J., Tremaine S.,  1987, Galactic Dynamics (Princeton:
Princeton Uviversity Press)}

{\bibitem{10} Brodie J. P., Schroder L. L., Huchra J. P., Phillips A. C.,
Kissler-Patig M., Forbes D. A., 1998, AJ, 116, 691}


{\bibitem{11} Brown R. J. N., Forbes D. A., Kissler-Patig M., Brodie J.,
2000, MNRAS, {\it in press}}

{\bibitem{11a} Bruzual A. G.m, Charlot S., 1996, electronically
available see: Leitherer C., et al., 1996, PASP, 108, 996}

{\bibitem{12} Burstein D., Heiles C., 1984, ApJS, 54, 33}

{\bibitem{13} Casoli F., Dupraz C., Combes F., Kazes I., 1991, A\&A, 251,
32}

{\bibitem{14} Couture J., Harris W. E.,  Allwright J. W. B.,  1990,
ApJS, 73, 671}

{\bibitem{15} Durrell P. R., Harris W. E., Geisler D., Pudritz R. E., 1996,
AJ, 112, 972}

{\bibitem{16} Elmegreen B., 1999, astro-ph/9911157}

{\bibitem{17}Elson R. A. W., Santiago B. X., 1996, MNRAS, 280, 971}

{\bibitem{18} Faber S. M., Gallagher J., 1979, ARA\&A, 17, 135}

{\bibitem{19}Franx M.,  Illingworth G.,  Heckman T., 1989, AJ, 98, 538}


{\bibitem{20} Forbes D. A., Grillmair C. J., Williger G. M., Elson R. A. W.,
Brodie J. P., 1998, MNRAS, 293, 325}

{\bibitem{21} Forbes D. A., Ponman T. J., Brown R. J. N., 1998, 508, L43}

{\bibitem{22} Forbes D. A., Brodie J. P., Grillmair C. J.,  1997, AJ, 113,
1652} 

{\bibitem{23} Forbes D. A., Hau G., 2000, MNRAS, in press}

{\bibitem{24} Georgakakis A., Forbes A. D., Norris P. R., 2000, MNRAS, in press}

{\bibitem{25} Gonzalez J. J., Faber S. M., Worthey G., 1993, AAS, 183, 4206}

{\bibitem{26} Goudfrooij P., Hansen L., Jorgensen H. E., Norgaard-Nielsen
H. U., de Jong T.,  van den Hoek L. B., 1994, A\&AS, 104, 179} 

{\bibitem{27} Harris W. E., 1986, AJ, 91, 822}

{\bibitem{28} Harris W. E., 1991, ARA\&A, 29, 543}

{\bibitem{29} Harris W. E., 1996, AJ, 112, 1487}

{\bibitem{30} Harris W. E., Harris G. L. H., McLaughlin D. E., 1998, AJ, 115,
1801} 

{\bibitem{31} Hernquist L., ApJ, 1992, 400, 460}

{\bibitem{32} Hibbard J. E., van Gorkom J. H., AJ, 1996, 111, 655}

{\bibitem{33} Kauffmann G., Charlot S., 1998, MNRAS, 294, 705}

{\bibitem{34} Kissler-Patig M., Brodie J. P.,  Schroder L. L., 
Forbes D. A., Grillmair C. J., Huchra J., 1998, AJ, 115, 105}

{\bibitem{35} Landolt A. U., 1992, PASP, 104, 336}

{\bibitem{36} Lauer T. R.,  1985, ApJ, 292, 104}

{\bibitem{37} Lilly S. J., Cowie L. L., Gardner J. P., 1991, ApJ, 369, 79}

{\bibitem{38} Lutz D., 1991, A\&A, 245, 31}

{\bibitem{39} McLaughlin D. E., 1999, AJ, 117, 2398}

{\bibitem{40} Metcalfe N., Shanks T.,  Fong R., Jones L. R., 1991, MNRAS,
249, 498}  

{\bibitem{41} Metcalfe N., Shanks T., Fong R., Roche N., 1995, MNRAS, 274,
257}   

{\bibitem{42} Meurer G. R., Heckman T. M., Leitherer C., Kinney A., Robert
C., Garnett D. R., 1995, AJ, 110, 2665}

{\bibitem{43} Miller B. W., Whitmore B. C., Schweizer F., Fall S. M., 1997,
AJ, 114, 2381}

{\bibitem{44}  Mihos J. C., Hernquist L., 1996, ApJ, 464, 641}

{\bibitem{45} Oke J. B., Cohen J. G., Carr M., Cromer J., Dingizian A.,
Harris F., H. Labrecque S.,  Lucinio R., Schaal W., Epps H., Miller J.,
1995, PASP, 107, 375}  

{\bibitem{46} Peletier R. F., Davies R. L., Illingworth G. D., Davis L. E.,
Cawson M., 1990, AJ, 100, 1091} 

{\bibitem{47} Savage B. D. \& Mathis J. S.,  1979, ARA\&A., 17, 73}

{\bibitem{48} Schweizer F., 1987, in Nearly Normal Galaxies, ed. S. Faber
(Springer: New York), 18}

{\bibitem{49} Schweizer F., 1996, AJ, 111, 109}

{\bibitem{50} Schweizer F., Seitzer P., 1992, AJ, 104, 1039}

{\bibitem{51} Schweizer F., Seitzer P., Faber S. M., Burstein D., Ore
C. M. D., Gonzalez J. J., 1990, ApJ, 364, L33}

{\bibitem{52} Terlevich I. A., Forbes A. D., 2000, MNRAS, submitted}

{\bibitem{53} Tragger S. C., 1997, PhD Thesis}   

{\bibitem{54} Toomre A., Toomre J., 1972, ApJ, 178, 623}

{\bibitem{55} Toomre A., 1977, `The Evolution of Galaxies and Stellar
Populations', edited by B. M. Tinsley and R. B. Larson (Yale University,
New Haven), p. 401}

{\bibitem{56} van den Bergh S., 1995, Nature, 374, 215}

{\bibitem{57} van den Bergh S., 1994, AJ, 108, 2145}

{\bibitem{58} van den Bergh S., 1984, PASP, 96, 329}

{\bibitem{58a} Vazdekis A., Casuso E., Peletier R. F., Beckman J. E.,
1996, ApJS, 106, 307} 

{\bibitem{59} Wiklind T., Combes F., Henkel C., 1995, A\&A, 297, 643}

{\bibitem{60} Woods D., Fahlman G. G., 1997, ApJ, 490, 11} 

{\bibitem{61} Worthey G., 1994, ApJS, 95, 107}

{\bibitem{62} Whitmore B. C., Miller B. W.,
 Schweizer F., Fall S. M., 1997, AJ, 114. 1797}

{\bibitem{63} Whitmore B. C., Schweizer F., 1995, AJ, 109, 960}

{\bibitem{64} Whitmore B. C., Schweizer F., Leitherer C., Borne K., Robert
C., 1993, AJ, 106, 1354}

{\bibitem{65}Zepf S. E., Ashman K. M., English J., Freeman K. C., Sharples
R. M., 1999, AJ, 1999, 118, 752}

\end{thebibliography}
\end{document}